\documentclass[a4paper]{article}
\usepackage{latexsym}
\begin {document}

\title {\bf On the Fock Transformation in Nonlinear Relativity }
\author{A.~Bouda\footnote{Electronic address:
{\tt bouda\_a@yahoo.fr}}
\ and T.~Foughali \footnote{Electronic address:
{\tt fougto\_74@yahoo.fr }}\\
Laboratoire de Physique Th\'eorique, Universit\'e de B\'eja\"\i a,\\
Campus de Targa Ouazemour, 06000 B\'eja\"\i a, Algeria\\}

\date{\today}

\maketitle

\begin{abstract}
\noindent
In this paper, we propose a new deformed Poisson brackets which leads to the Fock coordinate
transformation by using an analogous procedure as in Deformed Special Relativity. We therefore
derive the corresponding momentum transformation which is revealed to be different from previous
results. Contrary to the earlier version of Fock's nonlinear relativity for which plane waves
cannot be described, our resulting algebra keeps invariant for any coordinate and momentum
transformations the four dimensional contraction $p_{\mu} x^{\mu} $, allowing therefore to
associate plane waves for free particles. As in Deformed Special Relativity, we also
derive a canonical transformation with which the new coordinates and momentum satisfy the usual
Poisson brackets and therefore transform like the usual Lorentz vectors. Finally,
we establish the dispersion relation for Fock's nonlinear relativity.
\end{abstract}

\vskip\baselineskip

\noindent
PACS: 03.30.+p

\noindent
Key words: Fock's transformation, Poisson's brackets, Momentum transformation, Energy-momentum vector.

\newpage

\section{Introduction}

Nonlinear relativity was developed in the two following distinct ways.
\begin{enumerate}

\item The first was the Fock coordinate transformation \cite{Fock}, which was built on a ratio of linear functions.
At large scale in time, it induces a time-varying speed of light \cite{Manida} but keeps invariant a length which
represents the universe radius. However, this formulation was rejected since it does not allow to describe coherently
plane waves for free particles.

\item The second was the Deformed Special Relativity (DSR) which keeps invariant, in addition to the speed of light,
a minimal length in the order of the Planck length \cite{Amel-Pir, Ameliano1, Ameliano2, Mag-Smol1, Mag-Smol2, KMM}.
Later it is proven that the resulting transformation can be reproduced in the context of a non commutative space-time
by modifying the Poisson brackets \cite{Ghosh-Pal}. However, this theory has a particular feature concerning the
dependence of the coordinate transformation law on the impulsion and energy of the test particle. As suggested in
\cite{Harikumar}, this will induce a dependence on the test particle properties of the electromagnetic field in
the high energy domain. This intriguing feature was also discussed in the context of the links established between
gravitation and electromagnetism \cite{Bo-Bel}.

\end{enumerate}

We indicate that these two theories of nonlinear relativity was developed for completely different motivations.
In this paper, we propose a new deformed Poisson brackets which depend upon a parameter $R$ with a dimension of a
length and which is revealed to be the one introduced in the Fock transformation, to within a minus sign. As suggested in
\cite{Manida}, this parameter may represent the radius of the visible part of the universe. By using an analogous procedure
as in DSR, we will show that  these Poisson brackets lead to the Fock coordinate transformation, which leaves $R$ invariant.
We therefore derive the corresponding momentums transformation and we will see that contrary to what it is asserted in
\cite{Manida}, this transformation is different from the Lorentz one unless we take the limit $R\rightarrow\infty$.
Furthermore, our resulting algebra keeps invariant for any coordinates and momentum transformations between inertial frames
the four dimensional product $ p_{\mu} x^{\mu}$. This allows therefore to associate plane waves for free particles, what is not
the case in the previous version of Fock's nonlinear relativity \cite{Manida, KMM}. As in DSR, we establish a canonical
transformation with which the new variables obey to the usual Poisson brackets and therefore transform like the components
of the usual Lorentz vectors.

The paper is organized as follows. In section 2, the new deformed Poisson brackets and the resulting deformed Lorentz algebra
are presented. In section 3, the Fock coordinate transformation is reproduced and the corresponding momentums transformation is
established. In section 4, the canonical transformation and the dispersion relation for Fock's nonlinear relativity are derived.
Section 5 is devoted to conclusion.

\section{The new deformed Poisson brackets}

On the basis of the following deformed Poisson brackets ($\mu,\nu,...=0,1,2,3,$ and $i,j,...=1,2,3$)
\begin {eqnarray}
    \{x^{i},x^{0}\} & = &  { x^{i} \over \kappa }, \\
  \{x^{i},x^{j}\}   & = &  0 , \\
   \{x^{i},p^{j}\}  & = & -\eta^{ij} , \\
\{p^{\mu},p^{\nu}\} & = & 0 , \\
    \{x^{0},p^{i}\} & = & {p^{i} \over \kappa } , \\
    \{x^{i},p^{0}\} & = &  0 , \\
    \{x^{0},p^{0}\} & = & -1+ { p^{0} \over \kappa } ,
\end {eqnarray}
which can be rewritten in the form
\begin {eqnarray}
\{x^{\mu},x^{\nu}\} &  = &  {1 \over \kappa}\left(x^{\mu}\eta^{0\nu}-x^{\nu}\eta^{\mu0}\right), \\
\{p^{\mu},p^{\nu}\} &  = &  0, \\
\{x^{\mu},p^{\nu}\} &  = &  -\eta^{\mu\nu}+{1 \over \kappa}\eta^{\mu 0}p^{\nu},
\end {eqnarray}
where $\eta^{\mu\nu}=(+1,-1,-1,-1)$ and $\kappa$ is a parameter identified as the ratio between the Planck
energy and the light speed, Ghosh and Pal \cite{Ghosh-Pal} reproduced the following momentum transformation
\begin {equation}
E'={\gamma \left(E-up_{x} \right)\over \alpha_{\kappa}}, \hskip6mm p'_{x}={\gamma \left(p_{x}-uE/c^{2}\right)
\over \alpha_{\kappa}},
\hskip6mm p'_{y}={p_{y} \over \alpha_{\kappa}}, \hskip6mm p'_{z}={p_{z} \over \alpha_{\kappa}}
\end {equation}
and coordinate transformation
\begin {equation}
t'=\alpha_{\kappa} \gamma \left( t-ux/c^{2} \right), \hskip6mm x'=\alpha_{\kappa} \gamma \left(x-ut\right) ,
\hskip6mm y'=\alpha_{\kappa} y, \hskip6mm z'= \alpha_{\kappa} z
\end {equation}
already derived in \cite{Mag-Smol1, Mag-Smol2, KMM} from other considerations. Here $u$ is the
translation velocity of the primed frame origin on the $x$ axis of the unprimed one,
$\gamma=(1-u^{2}/c^{2})^{-1/2}$ and $\alpha_{\kappa}=1+[(\gamma-1)E-\gamma up_{x}]/c\kappa$.
It is easy to check that the energy $E_{P}=c\kappa$, identified as the Planck energy, is an invariant
as it is also the case for the Planck length $l_{p}=\hbar c/E_{P}=\hbar/\kappa$. We observe that in the limit
$\kappa\rightarrow\infty$, the above transformations reduce to the Lorentz ones for the four-momentum and
four-coordinate vectors.

Since $\alpha_{\kappa}$ depends on $E$ and $p_{x}$, we observe that we can not use the notion of event as in
special relativity when we deal for example with the transformation coordinates of rod ends. In fact, we are
forced to associate to any event a test particle with energy and momentum. Furthermore, the formulation of the
electromagnetism within the framework of the DSR will induce a dependence on the test particle properties of the
electromagnetic field in the high energy domain \cite{Harikumar}.

Contrary to the DSR, the Fock coordinate transformation does not depend on the impulsion and energy
of the test particle. In the next section, we will derive it from the following new deformed Poisson brackets
\begin {eqnarray}
\{x^{\mu}  ,x^{\nu}\} & = & 0, \\
      \{x^{i},p^{j}\} & = & -\eta^{ij},   \\
     \{x^{i} ,p^{0}\} & = & {x^{i} \over R} , \\
      \{p^{i},p^{j}\} & = & 0,   \\
      \{p^{i},p^{0}\} & = & -{p^{i} \over R}, \\
      \{x^{0},p^{i}\} & = & 0 ,   \\
      \{x^{0},p^{0}\} & = & -1 + {x^{0} \over R},
\end {eqnarray}
which can be rewritten in the form
\begin {eqnarray}
\{x^{\mu},x^{\nu}\} & = & 0 , \\
\{x^{\mu},p^{\nu}\} & = & -\eta^{\mu\nu}+{1 \over R}\eta^{0 \nu }x^{\mu}, \\
\{p^{\mu},p^{\nu}\} & = & -{1 \over R}\left(p^{\mu}\eta^{0\nu}-p^{\nu}\eta^{\mu 0}\right),
\end {eqnarray}
where $R$ is a parameter which will be identified later. Note that the above relations reduce to the
usual ones if we take the limit $R \rightarrow \infty$. From these relations, it
is easy to check that the following Jacobi identities
\begin {eqnarray}
\{x^{\mu }, \{x^{\nu},x^{\lambda }\}\}+\{x^{\lambda }, \{x^{\mu },x^{\nu  }\}\}+\{x^{\nu}, \{x^{\lambda },x^{\mu  }\}\} & = & 0,  \\
\{p^{\mu},\{x^{\nu} ,x^{\lambda}\}\}+\{x^{\lambda },\{p^{\mu } ,x^{\nu}\}\}+\{x^{\nu },\{x^{\lambda} ,p^{\mu}\}\}       & = & 0,   \\
\{p^{\mu},\{p^{\nu},x^{\lambda}\}\}+\{x^{\lambda },\{p^{\mu },p^{\nu}\}\}+\{p^{\nu },\{x^{\lambda},p^{\mu }\}\}         & = & 0,   \\
\{p^{\mu},\{p^{\nu},p^{\lambda}\}\}+\{p^{\lambda },\{p^{\mu },p^{\nu}\}\}+\{p^{\nu },\{p^{\lambda},p^{\mu }\}\}         & = & 0,
\end {eqnarray}
are satisfied. The angular momentum is defined as
\begin {equation}
J_{\mu\nu}=x_{\mu}p_{\nu}-x_{\nu}p_{\mu}.
\end {equation}
As in DSR \cite{Ghosh-Pal}, by using the usual Poisson bracket properties, one can show that the transformation of
$x^{\lambda}$ and  $p^{\lambda}$ are modified
\begin {equation}
\{J_{\mu\nu},x_{\lambda}\}= \eta_{\nu \lambda} x_{\mu}-\eta_{\mu \lambda} x_{\nu} -
{x_{\lambda} \over R}  \left( x_{\mu} \eta_{\nu 0} -x_{\nu} \eta_{\mu 0} \right),
\end {equation}
\begin {equation}
\{J_{\mu\nu},p_{\lambda}\}= \eta_{\nu \lambda} p_{\mu}-\eta_{\mu \lambda} p_{\nu} +
{p_{\lambda} \over R}  \left( x_{\mu} \eta_{\nu 0} -x_{\nu} \eta_{\mu 0} \right),
\end {equation}
but the Lorentz algebra remains unchanged
\begin {equation}
\{J_{\mu\nu},J_{\lambda \gamma}\}= -\eta_{\mu \gamma}J_{\lambda \nu} + \eta_{\nu \gamma  }J_{\lambda\mu}
                                    + \eta_{\nu  \lambda }J_{\mu \gamma} - \eta_{\mu \lambda }J_{\nu\gamma} .
\end {equation}
%
%

\section{ Coordinates and momentum transformations }

As in Lorentz transformation, we define an infinitesimal transformation of any function $O=O(x^{\mu},p^{\nu})$ by
\begin {equation}
\delta O= \{-{1\over2}\omega_{\mu\nu}J^{\mu\nu}, O\},
\end {equation}
where $\omega_{\mu\nu}$ are infinitesimal and antisymmetric transformation parameters. Applying (31) to
the coordinates and momentum components and taking into account (28) and (29), we obtain
\begin {equation}
\delta x^{\lambda}= -\omega^{\mu\lambda}x_{\mu} + {1 \over R} \omega^{\mu 0}x_{\mu}x^{\lambda},
\end {equation}
\begin {equation}
\delta p^{\lambda}= -\omega^{\mu\lambda}p_{\mu} - {1 \over R} \omega^{\mu 0}x_{\mu}p^{\lambda}.
\end {equation}
In what follows, we choose transformations for which the only non vanishing parameter is
$\omega^{01} = -\omega^{10} \equiv \delta\phi$. Therefore, by coming back to the conventional notations,
we deduce from (32) and (33)
\begin {eqnarray}
\delta t & =  &  \left(-{x \over c} +  {tx \over R} \right) \delta\phi, \\
\delta x & =  &  \left(-ct +  {x^{2} \over R} \right) \delta\phi ,  \\
\delta y & =  &    {x y \over R}  \delta\phi ,  \\
\delta z & =  &   {x z \over R}  \delta\phi ,
\end {eqnarray}
and
\begin {eqnarray}
\delta E     & = &  - \left(cp_{x} +  {x E \over R} \right) \delta\phi, \\
\delta p_{x} & = &  - \left( {E \over c } +  {x p_{x} \over R} \right) \delta\phi,  \\
\delta p_{y} & = &  - {x p_{y} \over R}  \delta\phi , \\
\delta p_{z} & = &  - {x p_{z} \over R}  \delta\phi .
\end {eqnarray}
We indicate here that $x^{0}=ct$ and $c$ is a constant with a dimension of a velocity and, as we will see later,
it represents in the limit $R\rightarrow \infty$ the light speed. By defining
\begin {equation}
\cosh \phi \equiv \gamma, \hskip15pt \sinh \phi \equiv \gamma { u \over c},
\end {equation}
where $\gamma=\left(1-u^{2}/c^{2} \right)^{-1/2}$, and following the same procedure developed in \cite{Ghosh-Pal},
we obtain from (34)-(37)
\begin {equation}
t'={ \gamma \left( t-ux/c^{2} \right) \over \alpha_{R}}, \hskip6mm x'= {\gamma \left(x-ut\right) \over \alpha_{R}},
\hskip6mm y'={ y  \over \alpha_{R}}, \hskip6mm z'= {z  \over \alpha_{R}}
\end {equation}
and from and (38)-(41)
\begin {equation}
E' = \alpha_{R} \gamma \left(E-up_{x} \right) , \ \ p'_{x} = \alpha_{R} \gamma \left(p_{x}-uE/c^{2}\right) ,
\ \ p'_{y} = \alpha_{R} p_{y} , \ \ p'_{z} = \alpha_{R} p_{z} ,
\end {equation}
where
\begin {equation}
\alpha_{R}= 1 + { 1 \over R } \left[ (\gamma-1)ct - \gamma { u x \over c} \right].
\end {equation}
Of course, in the limit $R\rightarrow \infty$, (43) and (44) reduce to the Lorentz
transformations for the coordinates and the energy-momentum vector.

Relations (43) are the same to those obtained by Fock \cite{Fock} and other authors in \cite{Manida}, \cite{KMM}
and \cite{KKRY} with others methods except that the parameter $R$ used in the literature is different from the
ours by a minus sign. We stress to indicate that it is the ours which is positive. In fact, if we apply the first
relation in (43) and take $t=R/c$, we will see that $t'=t$, meaning that $R$ is an invariant, interpreted in the
literature as the radius of the visible part of the universe. However, if we use the previous transformation
law established in the literature, in order to show that $R$ is an invariant, it is necessary to take $t=-R/c$.

Concerning relations (44), it is the first time that this momentum transformation
is established. In \cite{Manida}, the author claimed that the momentum transforms like the
usual Lorentz vector. This will imply that the four-dimensional contraction
$p_{\mu}x^{\mu}=Et-\overrightarrow{p}\cdot\overrightarrow{x}$
is not an invariant, as indicated in \cite{KMM}. However, if we apply the present coordinate and momentum
transformations, (43) and (44), it is easy to show that the contraction $p_{\mu}x^{\mu}$ is an invariant.
This provide the possibility to describe correctly the plane waves in the context of Fock's nonlinear
relativity.

In order to construct the relativistic invariant in this context, let us remark that relation (45) and the first
one in (43) allow to write
\begin {equation}
 \alpha_{R}=\left(1-{ct\over R}\right) \left(1-{ct'\over R}\right)^{-1}.
\end {equation}
and that from (43) we can show that
\begin {equation}
\eta_{\mu\nu}x'^{\mu}x'^{\nu}= { 1 \over \alpha^{2}_{R}}  \eta_{\mu\nu}x^{\mu}x^{\nu}.
\end {equation}
Therefore, by using (46) in (47), we deduce that the expression
\begin {equation}
I_{x} \equiv \left( 1 - { ct \over R} \right)^{-2} \eta_{\mu\nu}x^{\mu}x^{\nu}
\end {equation}
is an invariant. In the same manner, with the use of (44), we obtain
\begin {equation}
\eta_{\mu\nu}p'^{\mu}p'^{\nu}=  \alpha^{2}_{R}  \eta_{\mu\nu}p^{\mu}p^{\nu}
\end {equation}
and by taking into account relation (46), it is easy to show that the following expression
\begin {equation}
I_{p} \equiv \left( 1 - { ct \over R} \right)^{2} \eta_{\mu\nu}p^{\mu}p^{\nu}
\end {equation}
is an invariant. We observe that if we take the limit $R\rightarrow\infty$, $I_{x}$ and $I_{p}$
reduce to the well-known invariants of special relativity.


\section{ Momentum and energy }

Relations (48) and (50) suggest to define canonical variables
\begin {equation}
X^{\mu} \equiv \left( 1 - {x^{0} \over R}  \right)^{-1} x^{\mu},
\end {equation}
and
\begin {equation}
P^{\mu} \equiv \left( 1 - {x^{0} \over R}  \right) p^{\mu}
\end {equation}
in such a way that the invariants $I_{x}$ and $I_{p}$ would be expressed as in special relativity
\begin {equation}
I_{x} = \eta_{\mu\nu}X^{\mu}X^{\nu}, \ \ \ \ \ \ \ \   I_{p} = \eta_{\mu\nu}P^{\mu}P^{\nu}.
\end {equation}
Contrary to (52), relation (51) is already known in the literature \cite{Manida, KKRY}.
It is interesting to observe that with the use of the usual Poisson bracket properties, we can
reproduce from (20), (21) and (22) the well-known relations
\begin {eqnarray}
\{X^{\mu},X^{\nu}\} & = & 0 , \\
\{X^{\mu},P^{\nu}\} & = & -\eta^{\mu\nu} , \\
\{P^{\mu},P^{\nu}\} & = & 0 ,
\end {eqnarray}
meaning that $X^{\mu}$ and $P^{\mu}$ transform like usual Lorentz vectors. It is easy to show that
the inverse transformations of (51) and (52) are
\begin {equation}
x^{\mu} = \left( 1 + {X^{0} \over R} \right)^{-1} X^{\mu},
\end {equation}
and
\begin {equation}
p^{\mu} = \left( 1 + {X^{0} \over R}  \right) P^{\mu}.
\end {equation}

In order to search for the expressions of the energy and momentum, let us derive the expression of
the infinitesimal invariant $ds$. As the canonical variables obey to the laws of special relativity,
we have
\begin {equation}
ds^{2} = c^{2}dT^{2}-d\overrightarrow{X}^{2},
\end {equation}
where $cT=X^{0}$ and $\overrightarrow{X}$ is the three-dimensional position vector
in the canonical coordinate space. Using (51) in (59), we can show that
\begin {equation}
ds= {c \over (1-ct/R)^{2}} \sqrt{1-{1\over c^{2}}\left[ \left
                      (1-{ct \over R} \right)\overrightarrow{v}+ {c \overrightarrow{x} \over R} \right]^{2}} dt,
\end {equation}
where $\overrightarrow{x}$ is the three-dimensional position vector and
$\overrightarrow{v} \equiv d\overrightarrow{x}/dt$ the velocity. From this expression,
we can deduce the speed of light by imposing the condition $ds=0$. It is clear that it
depends on $t$ and $\overrightarrow{x}$ but it goes to $c$ when we take the limit $R \rightarrow \infty$.

As in special relativity, with the use of canonical variables, the four-dimensional momentum can
be written as
\begin {equation}
P^{\mu} = mc {dX^{\mu}\over ds},
\end {equation}
where $m$ is the rest mass. Substituting (51) and (52) in (61), we get to
\begin {equation}
p^{\mu} = {mc \over (1-ct/R)^{2}} {dx^{\mu}\over ds} + {mc^{2}x^{\mu} \over R(1-ct/R)^{3}} {dt\over ds}.
\end {equation}
With the use of (60), we obtain
\begin {equation}
p^{\mu} = {m \over \sqrt{1-{1\over c^{2}}\left[ \left(
                      1-{ct \over R} \right)\overrightarrow{v}+ {c \overrightarrow{x} \over R} \right]^{2}}}
                      \left[ {dx^{\mu} \over dt} + { cx^{\mu}\over R \left(1-{ct \over R} \right)}
                      \right].
\end {equation}
It follows that the three-dimensional momentum and the energy are given by
\begin {equation}
\overrightarrow{p} = {m \over \sqrt{1-{1\over c^{2}}\left[ \left(
                      1-{ct \over R} \right)\overrightarrow{v}+ {c \overrightarrow{x} \over R} \right]^{2}}}
                      \left[ \overrightarrow{v}  + { c\overrightarrow{x}\over R \left(1-{ct \over R} \right)}
                      \right]
\end {equation}
and
\begin {equation}
E = cp^{0} = {mc^{2} \over \left( 1-{ct \over R} \right) \sqrt{1-{1\over c^{2}}\left[ \left(
                      1-{ct \over R} \right)\overrightarrow{v}+ {c \overrightarrow{x} \over R} \right]^{2}}}
\end {equation}
In expression (64), the presence of $\overrightarrow{x}$ in the additive term to the velocity
reminds us of the role of the potential vector in the momentum of a charged particle under the action of an
electromagnetic fields. As $\overrightarrow{p}$ depends on $t$ and $\overrightarrow{x}$, relation (64) indicates
that at large scale in space, with distances of the order of $R$, the space is not homogeneous and the momentum
is not conserved even for free systems. The above expression of the energy indicates also that, at large scale
in time, in the order of $R/c$, the time is not uniform and the energy is not conserved. However, in the limit
$R \rightarrow \infty$, we recover in (64) and (65) the well-known expressions of special relativity and then
the space homogeneity and the time uniformity.

In order to establish the new dispersion relation, let us define a parameter $M$ with a dimension of mass by
\begin {equation}
M^{2}c^{2} \equiv I_{p}
\end {equation}
From (50), the particle energy is
\begin {equation}
E = c \left[ { M^{2}c^{2} \over \left( 1 - ct / R \right)^{2} }+ p^{2} \right]^{1/2},
\end {equation}
where $p \equiv |\overrightarrow{p}|$ is the modulus of the three-dimensional momentum.
As the canonical variables transform like usual Lorentz vectors, it is reasonable to define the rest mass, $m$,
in terms of these variables by using the definition which works in special relativity \cite{GN}
\begin {equation}
{ 1 \over m }= lim_{P\rightarrow 0} { c \over  P } { \partial P^{0} \over \partial P },
\end {equation}
where $P \equiv |\overrightarrow{P}|$. By using (52), we obtain
\begin {equation}
{ 1 \over m }= { 1 \over 1-ct/R } lim_{p\rightarrow 0} { 1 \over  p } { \partial E \over \partial p },
\end {equation}
Taking into account expression (67) of the energy, (69) leads to
\begin {equation}
m = M.
\end {equation}
Thus, the parameter $M$ introduced in (66) represents the rest mass. Finally, by using relation (67), the
dispersion relation in Fock's nonlinear relativity is
\begin {equation}
E^{2} =  { m^{2}c^{4} \over \left( 1 - ct / R \right)^{2} }+ p^{2} c^{2}.
\end {equation}
We note that this relation can be confirmed if we use expressions (64) and (65) of $\overrightarrow{p}$ and $E$.
We observe also that in the limit $R \rightarrow \infty$, this relation reduces to the one of special relativity.


\section{ Conclusion}

In this paper, we proposed a new deformed Poisson brackets which allowed to reproduce the Fock
coordinate transformation. We then derived a new momentum transformation with which the contraction
$p_{\mu} x^{\mu} $ becomes an invariant and therefore useful for the description of the plane waves.
As in DSR, we also established a transformation with which the new coordinates and momentum,
called canonical variables, transform like usual Lorentz vectors. This allowed us to determine
the expressions of the energy and momentum in which the vector position $\overrightarrow{x}$ plays
an analogous role as the potential vector in the momentum of a charged particle moving under the action
of an electromagnetic field. We also derived the dispersion relation in the context of Fock's nonlinear
relativity. The presence of $t$ and $\overrightarrow{x}$ in the expressions of the energy and momentum
indicate that, at large scale in time and space, the time is not uniform and the space is not
homogenous. At small scale, we may take the limit $R \rightarrow \infty$ and therefore recover the
uniformity of the time, the homogeneity of the space and all the known results of special relativity.
We stress to indicate that the parameter $R$ introduced here differs from the one of the literature
\cite{Fock,Manida,KMM,KKRY} by a minus sign. We showed here that it is the ours which is positive.

\bigskip
\bigskip

\noindent
{\bf REFERENCES}
\vskip\baselineskip

\end {document}